\documentstyle[prd,aps,epsf,graphicx,epsfig]{revtex}

\DeclareFontFamily{OT1}{rsfs10}{}
\DeclareFontShape{OT1}{rsfs10}{m}{n}{ <-> rsfs10 }{}
\DeclareMathAlphabet{\mathscript}{OT1}{rsfs10}{m}{n}

\newcommand{\be}{\begin{equation}}
\newcommand{\ee}{\end{equation}}
\newcommand{\nn}{\nonumber}
\newcommand{\bea}{\begin{eqnarray}}
\newcommand{\eea}{\end{eqnarray}}
\newcommand{\ba}{\begin{array}}
\newcommand{\ea}{\end{array}}

\newcommand{\ns}{\normalsize}
\newcommand{\pt}{\partial}

\newcommand{\eqref}[1]{(\ref{#1})}

\def\a{\alpha}
\def\b{\beta}

\def\c{\chi}
\def\d{\delta}
\def\e{\epsilon}

\def\f{\phi}
\def\z{\psi}

\def\k{\kappa}

\def\m{\mu}
\def\n{\nu}

\def\p{\pi}

\def\r{\rho}
\def\s{\sigma}

\def\t{\tau}

\def\z{\zeta}

\def\G{\Gamma}

\def\bal{{\mbox{\boldmath $\alpha$}}}

\begin{document}

%%%%%%%%%%%%%%%%%%%%%%%%%%%%%%%%%%%%%%%%%%%%%%%%%%%%%%%%%%%%%%%%%%%%%%
\begin{titlepage}

\title{
\hfill{\ns SUSX-TH/02-001\\}
\hfill{\ns OUTP-01-60P\\}
\hfill{\ns hep-th/0201040\\[2cm]}
{\huge Baryogenesis by Brane-Collision}\\[1cm]}
\setcounter{footnote}{0}
\author{{\ns\large
          Mar Bastero-Gil$^1$\footnote{email: mbg20@pact.cpes.susx.ac.uk},
\setcounter{footnote}{3}
 Edmund J.~Copeland$^1$\footnote{email: e.j.copeland@sussex.ac.uk}~,
 James Gray$^2$\footnote{email: J.A.Gray2@newcastle.ac.uk},\\
\setcounter{footnote}{7}
 Andr\'e Lukas$^1$\footnote{email: a.lukas@sussex.ac.uk}and
 Michael Pl\"umacher$^3$\footnote{email: pluemi@thphys.ox.ac.uk}\\[1cm]}
      {\ns $^1$Centre for Theoretical Physics,
     University of Sussex}\\
      {\ns Falmer, Brighton BN1 9QJ, United Kingdom}\\[0.3cm]
      {\ns $^2$Department of Physics, University of Newcastle upon Tyne}\\
      {\ns Herschel building, Newcastle upon Tyne NE1 7RU,
      United Kingdom}\\[0.3cm]
      {\ns $^3$Department of Physics, Theoretical Physics, 
      University of Oxford}\\[-0.2em]
      {\ns 1 Keble Road, Oxford OX1 3NP, United Kingdom}}

%\date{}

\maketitle

\vspace{1cm}

\begin{abstract}
We present a new scenario for baryogenesis in the context of
heterotic brane-world models. The baryon asymmetry of
the universe is generated at a small-instanton phase transition
which is initiated by a moving brane colliding with the
observable boundary. We demonstrate, in the context of a simple
model, that reasonable values for the baryon asymmetry can be
obtained. As a byproduct we find a new class of moving-brane
cosmological solutions in the presence of a perfect fluid.
\end{abstract}

\thispagestyle{empty}

\end{titlepage}

\renewcommand{\thefootnote}{\arabic{footnote}}

%%%%%%%%%%%%%%%%%%%%%%%%%%%%%%%%%%%%%%%%%%%%%%%%%%%%%%%%%%%%%%%%%%%%%%%

\section{Introduction}

An important feature of brane-world models which has attracted some
attention~\cite{Dvali:1999pa} -- \cite{Copeland:2001zp} is the
possibility of branes moving in the course of the cosmological
evolution.  In this paper, we would like to propose and analyse a new
mechanism for creating the baryon asymmetry in the universe based on
moving-brane cosmology.

\vspace{0.4cm}

We will be working in the context of heterotic
M-theory~\cite{Horava:1996qa,Horava:1996ma,Witten:1996mz} which
constitutes the strong-coupling limit of $E_8\times E_8$ heterotic
string theory. More precisely, we will be interested in the
five-dimensional heterotic brane-world models obtained by
compactification on a Calabi-Yau
three-fold~\cite{Lukas:1999yy,Lukas:1999tt}. In these models,
five-dimensional space-time is bounded by two (3+1)-dimensional
boundary planes, carrying the observable and the hidden sector,
respectively. Additionally, these models may contain ``bulk''
three-branes, originating from M-theory five-branes, which can move
along the fifth, transverse
direction~\cite{Witten:1996mz,Lukas:1999hk}. It is the motion of these
bulk three-branes which we would like to use for our baryogenesis
scenario. Another crucial ingredient is the small-instanton
transition~\cite{Witten:1996gx,Ganor:1996mu} which occurs when the
bulk three-brane collides with one of the boundary planes. It then
gets ``absorbed'' by the boundary plane and, at the same time, the
properties of the four-dimensional theory on the affected plane is
changed. In particular, the gauge group and/or the number of
families can change due to the collision~\cite{Ovrut:2000qi}.
 
\vspace{0.4cm}

These small-instanton transitions constitute a type of phase transition with a property which is qualitatively new to cosmology. Normally the temperature of the gas of particles in the universe when a phase transition occurs is given by the mass scale associated with that transition. This is not the case for these brane collisions. The time at which the brane collides with an orbifold fixed point depends on kinematical factors relevant to the brane such as how fast it is moving and its initial position. However if the brane were to change the gauge group on the orbifold fixed point it hits into a smaller group then some of the gauge bosons would become heavy during the collision. The mass that these bosons would gain would be determined by the Calabi-Yau size - typically of order the GUT scale - not the temperature of the gas of particles on that fixed point at the time of collision. There are clearly many possible new cosmological scenarios that could be developed using such qualitatively new behavior. In this paper we shall restrict ourselves to giving one example of an exploitation of this phenomenon in order to give a detailed and focused analysis. We shall use this effect to develop a new scenario of baryogenesis.

\vspace{0.4cm}

Roughly, our mechanism for baryogenesis is then as follows.
We start with a state in the early universe where the expansion
is driven by a gas residing on the observable boundary plane
and the kinetic energy of the brane which moves towards the
observable brane. At this stage, the quasi massless spectrum on
the observable brane is given by an $N=1$ gauge theory with group
$G_{\rm SM}\times U(1)_{B-L}$, where
$G_{\rm SM}=SU(3)\times SU(2)\times U(1)$, three MSSM families of quarks
and leptons plus three right-handed neutrinos (RHNs) and their scalar 
partners. All these
particles are in relativistic equilibrium. When, eventually, the
three-brane collides with the observable boundary, the gauge group is
changed to $G_{\rm SM}$ due to the small-instanton transition
and, as a consequence, the RHNs become super-heavy. Their
out-of-equilibrium decay then generates a lepton asymmetry which,
via electroweak sphaleron processes is converted into a baryon
asymmetry in the conventional way. Our mechanism is, in some ways,
similar to a standard leptogenesis scenario 
\cite{Fukugita:1986hr}--\cite{review}. 
However, instead of a GUT
phase transition we are using a small-instanton phase
transition, a genuine string effect. As we will see later, there are
a number of other important differences, including a dependence of
the baryon asymmetry on the parameters of the small-instanton transition
and the decoupling of the transition temperature from the RHN mass.

\vspace{0.4cm}

The outline of the paper is as follows. In the next section, we
will explain our scenario in detail but on an informal level.
In section three, an explicit realisation of this scenario in terms
of a simple model is presented. The quantitative predictions
of this model for the baryon asymmetry are analysed in section
four and five. Section six contains a summary of our main results
and an outlook. Finally, in the appendix, we present a new class
of moving-brane cosmological solutions in the presence of a perfect
fluid which are relevant to our baryogenesis scenario and, possibly,
to a number of other applications, such as the inflationary
scenario of Ref.~\cite{Dvali:1999pa}.

%%%%%%%%%%%%%%%%%%%%%%%%%%%%%%%%%%%%%%%%%%%%%%%%%%%%%%%%%%%%%%%%%%%%%

\section{The scenario}

Before we explicitly describe our scenario let us briefly explain the
theoretical framework which we will be using. Throughout this paper,
we will be working in the context of Ho\v rava-Witten
theory~\cite{Horava:1996qa,Horava:1996ma,Witten:1996mz}, that is,
M-theory on the orbifold $S^1/Z_2$. The low-energy limit of this
theory is described by 11-dimensional supergravity coupled to two
$E_8$ super-Yang-Mills multiplets each residing on one of the
10-dimensional orbifold fixed planes. More specifically, we will be
dealing with the five-dimensional brane-world models that can be
obtained by compactifying this theory on Calabi-Yau
three-folds~\cite{Lukas:1999yy,Lukas:1999tt}. These models are
described by gauged five-dimensional $N=1$ supergravity theories in
the bulk coupled to $N=1$ gauge theories located on the now
four-dimensional orbifold planes. As usual, we will interpret one of
these orbifold planes as the observable sector and the other one as
the hidden sector.

In addition, M-theory five-branes can be included in the
compactification from 11 to five
dimensions~\cite{Witten:1996mz,Lukas:1999hk}. They wrap a
two-dimensional curve in the Calabi-Yau space, stretch across the four
uncompactified dimensions and are parallel to the orbifold
planes. Hence, they appear as three-branes in the five-dimensional
brane-world theory which are located somewhere in the bulk between the
two orbifold planes. Each of these three-branes carries an additional
$N=1$ supersymmetric theory. A crucial feature for our purpose is that
these three-branes are not fixed but, rather, can move along the
orbifold direction.

The specific form of the $N=1$ theory on, say, the observable orbifold
plane is determined by the details of the compactification, that is,
by the choice of Calabi-Yau manifold and internal vector bundle. The
internal vector bundle can be thought of as instantons on the
Calabi-Yau space which serve to break the $E_8$ gauge group to a
phenomenologically more favorable subgroup. For different such
instanton configurations one generally obtains different gauge groups
and different sets of matter fields on the orbifold plane.  It has
been shown that phenomenologically interesting low-energy theories can
be obtained in this
way~\cite{Andreas:1999ei}--\cite{Donagi:2000zs}. The dependence of the
low-energy spectrum on the internal instanton configuration will be of
particular importance for us.

A final theoretical ingredient which we need to discuss is the
small-instanton
transition~\cite{Witten:1996gx,Ganor:1996mu,Ovrut:2000qi}. This
process occurs when one of the three-branes in the five-dimensional
brane-world model collides with an orbifold fixed plane.  From a
five-dimensional viewpoint, this three-brane is then ``absorbed'' by
the orbifold plane and disappears from the brane-world theory. For a
more microscopical picture we recall that the three-brane originates
{}from an M 5-brane which wraps a curve in the Calabi-Yau space and,
hence, carries some internal structure. As the brane collides with the
orbifold plane, this structure is being converted into an $E_8$ gauge
instanton on the Calabi-Yau space. In other words, the internal
instanton configuration, associated with the orbifold plane in
question, is changed in such a collision. In accordance with the above
discussion, this generally also implies that the low-energy
gauge-group or the matter-field content on the affected orbifold plane
is altered~\cite{Ovrut:2000qi}. 

\vspace{0.4cm}

\begin{figure}[ht]\centering
        \includegraphics[height=12.5cm,width=17cm]{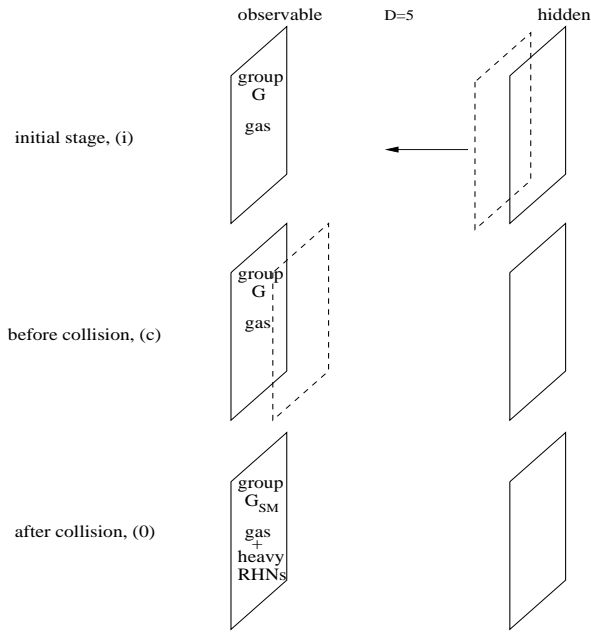} \vspace{-3.5cm}
        \caption{\emph{Shown are the three main stages of our
        scenario. The three-brane starts its evolution at some initial
        time $t_i$ when the observable plane carries a gas with
        temperature $T_i$ consisting of all degrees of freedom of an
        $N=1$ gauge theory with group $G\subset G_{\rm SM}$ plus
        matter fields (top). Shortly before the three-brane collides
        with the observable plane the gas has cooled to a temperature
        $T_c$ (middle). After the collision, the theory on the
        observable plane has been changed to the supersymmetric
        standard model due to the small-instanton phase transition.
        We now have a gas of standard model particles with temperature
        $T_0$ plus right-handed neutrinos which were massless in the
        original $G$ theory and have now become super-heavy
        (bottom). Their decay generates the lepton asymmetry.}}
        \label{fig1}
\end{figure}

We are now ready to discuss our baryogenesis scenario. We will be
working in the context of the five-dimensional brane-world model
described above, where we consider a single three-brane in the bulk,
for simplicity.  Let us consider a period in the early universe after
inflation where this three-brane moves along the
orbifold. We consider the existence of such a period a quasi-generic
feature of our brane-world model.  More precisely, we will analyse the
cosmological evolution of the model starting out from some initial
configuration, specified at time $t_i$.  At this time, we assume the
brane to be located at a specific point in the orbifold direction,
possibly close to the hidden orbifold plane, and having a certain
initial velocity pointing towards the observable orbifold plane.  In
addition, we assume that the energy density on the observable orbifold
plane is dominated by a gas with temperature $T_i$ while there is no
significant contribution to the energy density from the hidden
orbifold plane. One may, for example, think of this initial state as
the result of reheating after inflation.  At this stage, the
observable orbifold plane carries an $N=1$ gauge theory with gauge
group $G$ and a certain number of chiral matter fields. We assume that
this gauge group contains the standard model group $G_{\rm
SM}=SU(3)\times SU(2)\times U(1)_{Y}$, that is, $G_{\rm SM}\subset G$.
For concreteness, let us, in the following, discuss the "minimal
choice" $G=G_{\rm SM}\times U(1)_{B-L}$, where $B$ and $L$ are baryon
and lepton number, respectively. For simplicity, we also assume three
standard model families plus the same number of right-handed neutrinos
(RHNs) and their scalar partners. However, our mechanism will work 
for other groups $G$, such as
grand unified groups, and a more general matter field content as well
and the subsequent discussion can be easily modified accordingly. All
particles on the orbifold plane are exactly massless perturbatively
and it is plausible to assume that masses generated by
non-perturbative effects will not exceed the electroweak scale.  We
further assume that the temperature $T_i$ is much higher than that so
that the gas on the observable plane consists of all available
particles each being in relativistic equilibrium. We have, schematically,
depicted this initial state at the top of Fig.~\ref{fig1}.
\vspace{-0.3cm}

\vskip 0.4cm 

Starting from this initial state, the three-brane moves towards the
observable plane while the gas on this plane cools until it reaches
the temperature $T_c$ shortly before collision. This is shown in the
middle of fig.~\ref{fig1}. In the next section, we will describe this
evolution using the four-dimensional effective action associated with
our brane-world model. As we will see~\cite{Copeland:2001zp}, the
three-brane motion necessarily implies a time evolution of the
orbifold size as well as an evolution of the Calabi-Yau volume. We,
therefore, have to consider three scalar fields contributing to the
kinetic energy during this epoch.

When the three-brane finally collides with the observable plane, the
theory on this plane is changed. We assume that the initial $N=1$
gauge theory with group $G\subset G_{\rm SM}$ plus matter fields is
converted precisely into the supersymmetric standard model (MSSM) by
the small-instanton transition. Later on, we will present explicit
arguments that this can indeed be achieved for our concrete example
$G=G_{\rm SM}\times U(1)_{B-L}$. As a consequence, the right-handed
(s)neutrinos, previously effectively massless, now become super-heavy
with masses $M_i$, where $i$ is a generation index. What does this 
imply for the evolution of the gas on
the observable plane during the transition?
%Due to their $U(1)_{B-L}$ gauge interactions the RHNs and their scalar 
%partners are in relativistic equilibrium \cite{PluemiDipl} at 
%temperature $T_c$ before the collision. Similarly, all MSSM particles
%are in thermal equilibrium before the collision.
Before the collision, all MSSM particles are in thermal equilibrium
at temperature $T_c$. Further, due to their $U(1)_{B-L}$ gauge 
interactions, the RHNs and their scalar partners are also in relativistic 
equilibrium \cite{PluemiDipl}.
After the collision, we
have a gas of standard model particles in relativistic equilibrium at
temperature $T_0$ which we assume to be much larger than the
electroweak scale. It is plausible that a substantial number
of now super-heavy RHNs are still present after the collision. In
particular, if the characteristic time-scale of the collision is much
shorter than the RHN decay time, as we will assume later on, one
expects the number of RHNs to be basically conserved during the
transition. Their number density after the collision is then given by
the relativistic equilibrium distribution before collision. The decay
of those super-heavy RHNs then creates a lepton asymmetry. Since the
temperature $T_0$ is much higher than the electroweak scale, sphaleron
processes can then be invoked as usual to convert this into a baryon
asymmetry.

%%%%%%%%%%%%%%%%%%%%%%%%%%%%%%%%%%%%%%%%%%%%%%%%%%%%%%%%%%%%%%%%%%%%%%%%%%%%

\section{The model}

We would now like to be more explicit and realize our scenario in terms
of a simple model which will allow us to quantitatively estimate the
generated baryon asymmetry.

\vspace{0.4cm} 

First, we need to describe the three-brane motion across the orbifold
which we will do in terms of the relevant four-dimensional $N=1$
effective action. The minimal version of this action contains three
moduli fields, namely the dilaton $S$, the universal $T$-modulus and
the field $Z$ related to the position of the three-brane. The
K\"ahler potential for these fields is given
by~\cite{Derendinger:2001gy,Brandle:2001ts}
\begin{equation}
 K = -\ln\left( S+\bar{S}-q_5\frac{(Z+\bar{Z})^2}{T+\bar{T}}\right)
     -3\ln\left( T+\bar{T}\right)\; , \label{K}
\end{equation}
where $q_5$ is a constant. In terms of the underlying component fields,
these superfields can be written as~\cite{Brandle:2001ts}
\bea
 S &=& e^\f +q_5z^2e^\b -2i(\s -q_5 z^2\c) \nn \\
 T &=& e^\b +2i\c \label{STZ}\\
 Z &=& e^\b z-2i(\z -z\c )\nn\; .
\eea
Here the three real scalars $\s$, $\c$ and $\z$ are axionic fields,
which can be set to zero consistently. We will do this, for simplicity,
and work with the three remaining real scalar fields $\f$, $\b$ and $z$.
{}From the above expression for the K\"ahler potential, the action for
these component fields is given by
\begin{equation}
\label{cmp_action}
 S = 
-\frac{1}{2\k_P^2}\int\sqrt{-g}\left[\frac{1}{2}R+\frac{1}{4}\pt_\m\f
     \pt^\m\f +\frac{3}{4}\pt_\m\b\pt^\m\b+\frac{q_5}{2}e^{\b -\f}
     \pt_\m z\pt^\m z\right]\; . \label{S4}
\end{equation}
The interpretation of these scalar fields is as follows. The size
of the Calabi-Yau space and the orbifold are proportional to $e^\f$
and $e^\b$, respectively, while the position of the three-brane is
given by $z\in [0,1]$ where $z=0$ corresponds to the observable (say)
orbifold plane and $z=1$ to the hidden one. Perturbatively, these
fields represent flat directions but at non-perturbative level a potential
may have to be added to the above action. For simplicity, we will not
consider such a potential explicitly which amounts to assuming that
the energy density in the gas and the kinetic energy dominate the
potential energy. As mentioned earlier, a moving three-brane necessarily
implies time-evolution of the fields $\f$ and $\b$, as can be seen
{}from the kinetic term of the $z$ field in Eq.~\eqref{S4}.

\vspace{0.4cm}

The cosmological solutions to the action~\eqref{S4} with a moving
three-brane but without a gas have been found in
Ref.~\cite{Copeland:2001zp}.  Here, we will need the generalisation of
those solutions to include the stress energy due to a gas with
pressure $p_{\rm gas}=\r_{\rm gas} /3$ located on the observable brane
at $z=0$. Remarkably, these solutions can be found exactly even for
the more general equation of state
$p_{\rm fluid}=w\r_{\rm fluid}$ where $w$ is a
constant. They are explicitly given in Appendix~\ref{app}. We stress
that for our application to baryogenesis we will be using the
positive-time branch of those solutions. Hence, unlike the
inflationary scenario of Ref.~\cite{Khoury:2001wf} and pre-big-bang
cosmologies in general, our model has no exit problem. What we
need for our application is not so much the detailed form of the
solution but, rather, the relation between the initial data, provided
at time $t_i$ and the data before collision at time $t_c$.  Let us
define the ratio
\begin{equation}
 r(T) = \frac{\r_{\rm kin}}{\r_{\rm gas}}
\end{equation}
of the total kinetic energy of the three scalar fields and the
energy density of the gas. In terms of our explicit model, both
quantities are defined in Eq.~\eqref{kin} and \eqref{fluid}.
As we will see, for our simple description
of the collision later on, the quantity $r_c\equiv r(T_c)$ and the
temperature $T_c$ is all we really need to know as inputs right before
the collision. The question is then how these quantities depend on the
initial data and whether they are constrained in any way.
Clearly, for all our solutions the ratio $r(T)$ scales as
\begin{equation}
 \frac{T_i}{T_c}=\left[\frac{r_i}{r_c}\right]^{1/2}\; ,\label{r}
\end{equation}
where $r_i\equiv r(T_i)$. This relation, of course, simply reflects
the standard scaling properties of radiation and kinetic energy.  Of
course, there is no a priori information about the initial data
although there may be plausible assumptions about their nature~\footnote{An
example is to assume that $r_i\sim 1$, so that kinetic and radiation
energy density are of the same order, initially. Then, from Eq.~\eqref{r},
we have $r_c\ll 1$ for only a moderate decrease in temperature and, hence,
a case with small brane impact.}.
However, these data are constrained by requirements to be imposed at
time $t_c$ before collision. First of all, we need to pick a solution
where, at time $t_c$, the three-brane indeed collides with the
observable plane at $z=0$. In addition, we may require that the size
of the Calabi-Yau space and the orbifold at time $t_c$ are in the
appropriate range for gauge unification in the sense of
Ref.~\cite{Witten:1996mz}, thereby imposing constraints~\footnote{For
a fully realistic model, one would have to include non-perturbative
stabilising potentials for these fields.} on $\f (t_c)$
and $\b (t_c)$. One can demonstrate from the explicit form of the
solutions in App.~\ref{app} that all these constrains can be satisfied
simultaneously by choosing appropriate initial conditions and that, by
doing so, no further constraints on $T_i$, $r_i$, $T_c$ and $r_c$
other than Eq.~\eqref{r} are imposed. We will therefore use
Eq.~\eqref{r} as the single relation to link the initial state with
the state before collision.

\vspace{0.4cm}

We should now discuss the brane collision. First, how do we realize
the required transition of the gauge group from
$G=G_{\rm SM}\times U_{B-L}(1)$ to
$G_{\rm SM}$? To obtain $G$ we need an internal bundle
with structure group $SU(4)\times Z_n$ where $Z_n$ corresponds
to a Wilson line. The $SU(4)$ part serves to break the original
$E_8$ to $SO(10)$ and the $Z_n$ Wilson line can be chosen to break
$SO(10)$ precisely into $G$. To realize $G_{\rm SM}$ the required
bundle structure is $SU(5)\times Z_m$ with an appropriate $Z_m$
Wilson line. From a purely group-theoretical viewpoint, the
$SU(5)$-breaking Wilson line $Z_m$ can be chosen as the intersection
of the $SO(10)$-breaking Wilson line $Z_n$ with $SU(5)$. This
suggests that the Wilson lines can be viewed as ``spectators''
and that the required transition of the internal bundle is
basically $SU(4)\rightarrow SU(5)$. Such transitions
can indeed be obtained for suitable compactifications
and explicit example have been given in Ref.~\cite{Ovrut:2000qi}.

Our main task is now to determine the basic initial conditions for
leptogenesis, that is the initial number density of RHNs and the
temperature at the beginning of leptogenesis, in terms of the
parameters of our model. We have three generations of matter fields,
in particular three RHNs with associated heavy mass scales $M_i$,
where $i=1,2,3$. For simplicity, in this and the following section,
we will discuss the single-family case focusing on the first generation
with corresponding RHN mass $M=M_1$.  It should be noted that this mass is the mass of the RHN $\it{after}$ the brane collision. Our results are easily generalised and
all three flavours will be included in the numerical simulation, later on.
Unfortunately, a detailed microscopical
understanding of the dynamics of the small-instanton transition is
well beyond present knowledge and we will not attempt to improve on
this in the present paper. Instead we will rely on a
``phenomenological'' description mainly based on three simple
assumptions in order to analyse our scenario. First of all, we assume
continuity of the scale factor and its derivative across the
transition. This allows us to match the total energy densities before
and after. Secondly, we assume that the energy density after the
collision is dominated by the gas of standard model particles and
super-heavy RHNs. This asserts, among other things, that the scalars
$\f$ and $\b$ do not carry significant kinetic energy after the
collision~\footnote{The same is required for the bundle modulus
which corresponds to the size of the small instanton.}.
While a generalisation to include scalar field evolution
after collision may be feasible, we would like to focus on the rather
simpler case here.  With these two assumptions, the energy density
matching simply reads
\begin{equation}
 \r_{\rm gas}(T_c)+\r_{\rm kin}(T_c)=\r_{\rm gas}(T_0)+\r_N(T_0)\; ,
 \label{match}
\end{equation}
where $\r_N$ is the energy density in RHNs and their scalar partners. 
An additional constraint is given on this energy density matching by the second law of thermodynamics. This means in the present situation that we should not allow thermal energy to be converted into some more ordered form. Thus we impose the constraint $T_0>T_c$. Intuitively we would not expect to see situations where the energy from the five-brane position modulus is converted with perfect efficiency into RHN's; we would expect $T_0$ to be at least a bit larger than $T_c$. However it takes very little extra effort to consider cases where the two temperatures are practically the same and so we shall not unnecessarily restrict the possibilities which we consider in our subsequent analysis.  Our third and final assumption is
that the transition time is much shorter than the decay time of
the RHNs and, in fact, electroweak interaction rates. This implies
that number densities of all particles in the initial gas are
essentially unchanged across the transition. If the brane impact
is sufficiently large, so that a significant amount of energy is
transferred to the gas, one expects the gas to be out of equilibrium
after the transition. Given our ignorance about the details of the
transition, such a situation will be hard to describe quantitatively.
We, therefore, require that equilibrium is restored on a time-scale
much shorter than the RHN decay time. We will identify the third
stage of our scenario, corresponding to temperature $T_0$, with this
particular time when equilibrium has been restored after the transition.
Let us now analyse the conditions for such a swift thermalisation.
The typical ratio of RHN decay rate $\G_N$ and electroweak interaction
rate $\G_I$ is given by
\begin{equation} 
\frac{\Gamma_N}{\Gamma_I} \sim \frac{ |h_{\nu}|^2}{\alpha_{EW}} \frac{M}{T_0} \, ,
\label{Gr}
\end{equation}
where $h_{\nu}$ is the RHN Yukawa coupling and $\alpha_{EW}$ is
a typical standard model coupling. 
We should now distinguish the two cases
$T_0\gg M$ and $T_0\ll M$. In the former case the ratio~\eqref{Gr} is
suppressed by $\frac{M}{T_0}$ and all particles including the RHNs get into
equilibrium well before the RHNs decay. Hence, the RHNs are in
relativistic equilibrium at temperature $T_0$. This determines the initial
conditions for leptogenesis in the case $T_0\gg M$ which are similar
to the ones in the standard leptogenesis scenario. In particular,
due to the thermalisation of RHNs after the collision, the initial
number density of RHNs does not depend on the parameters of the
small-instanton transition.

\vspace{0.4cm}

On the other hand, if $T_0\ll M$, the RHNs are non-relativistic after
collision and will not return to equilibrium. In order for the gas of
MSSM particles to thermalise before the RHNs decay we should require,
from Eq.~\eqref{Gr},that $|h_{\nu}|^2 < \alpha_{EW}  \frac{T_0}{M}$. This puts a
mild constraint on the RHN Yukawa coupling which we will assume to be
satisfied in the following. Then, the RHN number density at $T_0$
is given by its equilibrium value before collision, that is, by
\begin{equation} \label{n0}
 n_N(T_0)=n_N(T_c)=\frac{3\z (3)}{4\p^2}g_NT_c^3\; , \label{nN}
\end{equation}
where $g_N$ is the number of degrees of freedom in the RHN supermultiplet. 
Further, we can use the standard expression 
\begin{equation}
 \r_{\rm gas}(T)=\frac{\p^2}{30}g_*(T)T^4
\end{equation}
for the energy density of a gas at temperature $T$. Applying
the energy matching condition~\eqref{match} to relate the temperatures
$T_c$ and $T_0$ before and after collision one finds~\footnote{We
recall, that we have assumed, for simplicity, that the number of
families is unchanged by the transition and, hence, $g_*(T_c)\simeq
g_*(T_0)$.  If we had allowed for a change in the number of
generations, the right-hand side of Eq.~\eqref{Tratio} would have to
be multiplied by $g_*(T_c)/g_*(T_0)$.}
\begin{equation}
 \left(\frac{T_0}{T_c}\right)^4 =  \left(1+r_c-\d \frac{M}{T_c} \right)\; .\label{Tratio}
\end{equation}
where we have defined the constant
\begin{equation}
 \d = \frac{45\z (3)g_N}{2\p^4g_*(T_c)}\approx 10^{-2}\; .
\label{gamdel}
\end{equation}
Eq.~\eqref{Tratio} represents the crucial matching condition for
the case $T_0\ll M$. The inequality $T_0\ll M$, expressed in
terms of the initial data at temperature $T_c$, takes the form
\begin{equation}
 \frac{M}{T_c}\gg (1+r_c)^{1/4}\; .\label{rel}
\end{equation}
The interpretation of the right-hand side of Eq.~\eqref{Tratio} in
terms of an energy balance is straightforward. The first two terms
represent the positive contributions from the energy density stored in
the gas before collision and the kinetic energy of the scalar
fields. Accordingly, the cases $r_c\ll 1$ and $r_c\gg 1$ correspond to
a collision with small and large impact respectively. The third term
is due to the RHNs becoming massive and, hence, it contributes with a
negative sign. Of course, we have to ensure that the energy density
before collision is sufficiently large to account for the masses of
the RHNs given the restriction on energy density redistribution imposed by the second law. In other words the right-hand side of
Eq.~\eqref{Tratio} must be greater than 1~\footnote{If this condition is violated
we would expect that our assumption of a fast phase transition would be
invalid.}. An even more stringent constraint results in some cases from the fact that there should be enough energy so
that the temperature after collision, $T_0$, exceeds a certain minimal
temperature even if $T_c$ does not. Since we would like sphaleron effects to convert the
lepton asymmetry into a baryon asymmetry this minimal temperature
should correspond to the electroweak scale $T_{\rm EW}$. This leads to
the following constraint
\begin{equation}
 \left(\frac{T_{\rm EW}}{T_c}\right)^4< \left(1+r_c- \frac{\d M}{T_c} \right)\; .\label{kl}
\end{equation}
Depending on the parameters, after the collision, either the gas or
the massive RHNs may dominate the energy density. It turns out that for
\begin{equation}
  T_c > M \left( \frac{2\d}{1+r_c} \right)\; .\label{mr}
\end{equation}
the universe is radiation dominated after collision, that is
$\r_N (T_0)<\r_{\rm gas}(T_0)$, and matter dominated otherwise.
For the case $T_0\ll M$, we are now ready to express two crucial input
quantities for leptogenesis in terms of the parameters of our model.
These are the temperature $T_0$ at which the RHNs start to decay and
the number of RHNs per entropy density $Y_N(T_0)$. We find
\bea
 T_0 &=& \left[ \left(1+r_c-\d \frac{M}{T_c} \right)\right]^{1/4}T_c \\
 Y_N(T_0) &=& \frac{3}{4}\d\left(\frac{T_c}{T_0}\right)^3
           =  \frac{3}{4}\d\left[1+r_c-\d \frac{M}{T_c}\right]^{-3/4}\; ,
\label{yn0}
\eea
where we recall from Eq.~\eqref{r} that
\begin{equation}
 r_c = \frac{T_c^2}{T_i^2}r_i\; .
\end{equation}
In the following section, we will use these input values for a simple
analytic estimate of the baryon asymmetry if $T_0\ll M$. Such an
estimate is possible since at these low temperatures lepton number
violating scattering processes mediated by the RHNs are inoperative.
For $T_0\sim M$ or $T_0\gg M$, on the other hand, the wash-out due
to scattering can be significant and a quantitative description requires
the numerical integration of the full set of Boltzmann equations.
This will be discussed in further detail in section five.

%%%%%%%%%%%%%%%%%%%%%%%%%%%%%%%%%%%%%%%%%%%%%%%%%%%%%%%%%%%%%%%%%%%%%%%%%%

We remark again that we are assuming the transition time to be shorter
than the decay time of the RHNs. This excludes the possibility of 
significantly decreasing the number density of the RHNs across the
transition. Nevertheless, it could still be possible to produce 
RHNs through  parametric resonance, due to
a non adiabatic change in their mass \cite{preheating}. Particle
production then takes place when the adiabaticity condition is
violated and $M^2 \ll |\dot M| $. Typically, this happens over a short
period of time $\delta t_*$, when the field dependent mass of the RHNs is very
small. In our scenario this would occur at the beginning of the
transition,  when the mass is changing from zero to
non-zero values; we do not expect further particle
production afterwards. The number density of particles
produced can be estimated as \cite{instant}
\be
n_N \sim \frac{ (\dot M(t_*))^{3/2}}{8 \pi^3} \,. \label{nNpreh}
\ee 
Comparing Eq.~(\ref{nNpreh}) with Eq.~(\ref{nN}), our assumption that
the number density for the RHNs remains unchanged through the transition
implies that
\be
\dot M(t_*) < \rho_{gas}(T_c)^{1/2}  \label{mdot}\, .
\ee
To evaluate this condition, we need information about the time
variation $\dot M(t_*)$ of the RHN mass during the collision.
Obviously, precise statements can only be made on the basis
of a microscopic description of the transition. For some more
qualitative information, however, we note that the mass
of the RHNs is correlated with the size modulus of the small instanton..
This modulus can, in turn, be interpreted as the ``continuation'' of
the field $z$. Hence, one expects $\dot M(t_*)$ to be small for
small $\dot z(t_*)$. For small impact parameter, $r_c < 1$, 
the parametric resonance will, therefore, not significatively
change the RHN number density. For large impact parameter, $r_c >1$, 
we would need to know the precise relation between $M(t)$ and $z(t)$,
that is, the details of the transition, to estimate the number density of the
particles produced. With these details being unavailable, we cannot
exclude the possibility of additional RHN production for large impact
parameters. From a conservative viewpoint, our subsequent results
for the baryonasymetry at large brane impact should, therefore, be
interpreted as lower bounds.

%%%%%%%%%%%%%%%%%%%%%%%%%%%%%%%%%%%%%%%%%%%%%%%%%%%%%%%%%%%%%%%%%%%%%%%%%%

\section{Estimate of the baryon asymmetry}
\label{est}

In this section we focus on the case $T_0<M$. The lepton asymmetry
is generated in the decays of the RHNs due to the CP asymmetry
$\e_i$ in the decay into (s)leptons and anti-(s)leptons
which arises due to interference between tree-level and one-loop
diagrams~\cite{Covi}:
\begin{eqnarray}
\epsilon_i 
    &=&-\frac{1}{8\pi(h_{\nu} h_{\nu}^\dagger)_{11}}
\sum_j \left( {\rm Im} \left[ (h_{\nu}
h_{\nu}^\dagger)_{ij}\right]^2 \right) f(M_j^2/M_i^2)\,,
\end{eqnarray}   
where $h_{\nu}$ are the Yukawa couplings and 
\begin{equation} \label{f}
f(x) = \sqrt{x} \left[ \log \left( \frac{1+x}{x} \right)+{2\over x-1}
\right] \,.
\end{equation}
The second term in Eq.~(\ref{f}) originates from the one-loop self-energy,
which can only be reliably calculated in perturbation theory for
sufficiently large mass splittings \cite{Buchmuller:1998yu}. 
In the case of small mass splittings
this contribution could be enhanced \cite{lgrev}. However, 
since Eq.~(\ref{f}) has
been calculated in standard zero temperature perturbation theory, the
influence of thermal effects on the self-energy term is unclear.
%but then the contributions of
%the three generations of RHNs to the lepton asymmetry will partially
%cancel each other, in analogy to the GIM mechanism in the standard model.
For simplicity we will therefore assume a hierarchy of the form
$M_1\ll M_2,M_3$~\footnote{In principle, all masses, or at
least Yukawa couplings, should be computable in a given M-theory model.
While it would be interesting to analyse this in more detail in relation
to our scenario, here we will simply treat those parameters as
phenomenological quantities.}. Of course, all three generations of
RHNs could still contribute to the lepton asymmetry. However, if one
assumes such a hierarchy there should not be strong cancellations
between the different contributions. As in the previous section, we will
focus on the first family, setting $\e_{\rm CP}=\e_1$.
Our result can then be easily generalised to apply to all flavours
$i$ with $M_i/T>1$. Eq.~(\ref{f}) also gives the CP asymmetry in the
decays of scalar right-handed
neutrinos into (s)leptons and anti-(s)leptons. At temperatures far
above the electroweak scale, where SUSY breaking effects can be neglected,
they will give the same contribution to the lepton asymmetry as the
RHNs \cite{PluemiDiss}. For simplicity, we will only mention RHNs 
in the following, but the contributions from their scalar partners 
will be included in our results.

$B+L$ violating sphaleron processes, which are
in equilibrium for temperatures between $O(10^{12})$ GeV and
$T_{EW}$, will partially convert the produced lepton asymmetry $Y_L$
into a baryon asymmetry $Y_B$, giving rise to the following relation
between baryon and lepton asymmetries:
%If there is no further dilution after the RHNs decay, at the
%time of nucleosynthesis, we have
\begin{equation} 
Y_{B} = -\frac{8}{23} Y_{L} \;.
\end{equation} 
%with $Y_{L}$ computed at the time of the decay. 
The observed value
for the baryon asymmetry is given by $Y_{B}\sim 10^{-10}$.

In the case $M> T_0$, considered in this section, decay processes
dominate over scattering processes. Hence, the scenario for
leptogenesis simplifies to that of the out-of-equilibrium decay of a
massive, non-relativistic species (RHNs) into light degrees of
freedom. The evolution of the system composed of heavy RHNs and a gas
of relativistic particles can be described by the equations
\cite{kolbturner}:
\begin{eqnarray} 
\dot \rho_N + 3 H \rho_N + \Gamma_N \rho_N &=& 0 \,, \label{rho} \\
\dot \rho_{\rm gas} + 4 H \rho_{\rm gas} - \Gamma_N \rho_{N} &=& 0 \,, \\
\dot n_{L} + 3 H n_{L} - \epsilon_{\rm CP} \Gamma_N
\frac{\rho_N}{M} \, &=& 0 \,, \label{nl} 
\end{eqnarray} 
where $n_{L}$ is the density of the net lepton number
generated in the decays, $\Gamma_N = (h_{\nu}h_{\nu}^\dagger)_{11} 
M_1/(4 \pi)$ is the
decay rate, and $H$ is the  Hubble rate, given by
$$
H^2= \frac{8 \pi}{3M_P^2} ( \rho_N(T) + \rho_{\rm gas}(T) ) \,.
$$
{}From Eqs. (\ref{rho}-\ref{nl}) the evolution of the number per
entropy densities $Y_N$, $Y_L$ can be written as
\begin{eqnarray}   
\dot Y_N &=& - Y_N (\Gamma_N + \frac{\dot S}{S}) \,, \label{yn}\\ 
\dot Y_{L} &=& \epsilon_{\rm CP} \Gamma_N Y_N - Y_{L}
\frac{\dot S}{S} \,, \label{yl} 
\end{eqnarray} 
where $S$ is the entropy per comoving volume.
The final value of the lepton asymmetry after the RHNs have decayed 
($t_f \gg \Gamma_N^{-1}$) is then given by
\begin{equation} 
Y_{L} = \epsilon_{\rm CP} Y_N(T_0) \frac{S_0}{S_f} \label{yls}\,,
\end{equation} 
where $Y_N(T_0)$ is given in Eq. (\ref{yn0}) in terms of the
parameters of the model. The factor $S_0/S_f$ is the usual 
dilution factor due
to entropy production during the decay. That is, assuming that the
decay products rapidly thermalise, they heat up the Universe and
contribute to the total entropy at a rate  
\begin{equation}
\dot S = \Gamma_N \frac{ a^3 \rho_N}{T} \,,
\end{equation}  
where $a$ is the scale factor. Entropy production will be significant
if the total energy density is dominated by the RHNs, that is,
when the decay products can make a non negligible contribution to the
thermal bath. On the other hand, if at $T_0$ the energy density was
dominated by the gas of relativistic particles, the decay products will
make almost no difference to the entropy. Therefore, the ratio $S_f/S_0$
is bounded between unity and the value \cite{turner} 
\begin{eqnarray} 
\frac{S_f}{S_0} &\simeq& \left(1+ \left(
\frac{\bar{g}_*}{g_*(T_0)}\right)^{1/3} 
\left(\frac{\rho_N(T_0)}{\rho_{\rm gas}(T_0)}\right)
\left(1+\frac{H(T_0)}{\Gamma_N} \right)^{2/3} \right)^{3/4} \nonumber\,, \\
& \simeq& \left(
\frac{\bar{g}_*}{g_*(T_0)}\right)^{1/4} 
\left(\frac{\rho_N(T_0)}{\rho_{\rm gas}(T_0)}\right)^{3/4}
\left(1+\frac{H(T_0)}{\Gamma_N} \right)^{1/2} \,, \label{entr}
\end{eqnarray} 
where $\bar{g}_*$ is the average number of light degrees of
freedom between $T_0$ and $T_f$, and 
\begin{equation}
\frac{H(T_0)}{\Gamma_N} = \frac{1}{2} 
\left(\frac{T_c}{T_f}\right)^2 \sqrt{ 1 +r_c} \,,
\end{equation}  
with $T_f \simeq 0.5 g_*(T_0)^{-1/4} \sqrt{M_P \Gamma_N}$.

\vspace{0.4cm}

Let us first discuss the case where the RHNs dominate the energy density
after collision, that is, when significant entropy is produced.
This corresponds to the parameter range
\begin{equation}
\frac{\d M}{r_c} <T_c<\frac{2\d M}{1+r_c} \label{l1}
\end{equation}
of our model, where we recall that $\d\approx 10^{-2}$ and $r_c\ll 1$
($r_c\gg 1$) corresponds to a small (large) impact collision.
We remind the reader that the upper bound in Eq.~\eqref{l1} guarantees
that the universe is matter-dominated after the collision,
see Eq.~\eqref{mr}, while the lower bound is the kinematic limit which ensures
that enough energy is available to account for the RHN mass while remembering that we must obey the second law of thermodynamics,
see Eq.~\eqref{kl} in the limit $T_{\rm EW}/T_c\ll 1$. Note that if $r_c$ is too small there is not enough energy available in this regime to account for the mass of the RHNs and the fact that the temperature of the gas cannot decrease during the collision. In other words if $r_c<1$ the parameter range given by \eqref{l1} closes up completely.

{}From Eqs.~\eqref{yls} and \eqref{entr} we then find a lepton asymmetry
of the order  
\begin{equation}  
Y_{L} \simeq \epsilon_{\rm CP} g_*^{-1/4} \frac{ \sqrt{M_P
\Gamma_N}}{M} \,, \label{ylmin} 
\label{ymin}
\end{equation} 
where the number of relativistic degrees of freedom
$g_*\simeq g_*(T_0)$ is taken to be practically constant through the
decay. Hence, for this case, the final baryon asymmetry does not
depend on the parameters of the small instanton transition. In particular,
it is independent of the initial number of RHNs given in Eq.~\eqref{yn0}.
For typical values $\e_{\rm CP}\sim 10^{-6}$ -- $10^{-8}$ one can clearly
obtain an acceptable value for baryon asymmetry, in this case.

\vspace{0.4cm}

On the other hand, if there is no significant entropy production and,
hence, the universe is radiation dominated after the collision,
we should consider the parameter range
\begin{equation}
 \frac{2 \d M}{(1+r_c)}<T_c<\frac{M}{(1+r_c)^{1/4}} \; ,
\end{equation}
for the case $r_c \geq1$, and
\begin{equation}
\frac{ \d M}{r_c}<T_c<\frac{M}{(1+r_c)^{1/4}} \; ,
\end{equation}
for $r_c<1$.
The upper bound is equivalent to $T_0<M$, implying non-relativistic
RHNs after the collision, see Eq.~\eqref{rel}, while the lower
bound guarantees a radiation-dominated universe after collision and that the energy matching requirements can be met,
see Eq.~\eqref{mr} for the first of these. Note that if the brane is not moving fast enough (if  $r_c< \d$) then the energy matching requirements can not be met in this regime either. Using Eqs.~\eqref{yls} and \eqref{yn0} the lepton
asymmetry is then given by
\begin{equation} 
Y_{L} \simeq \epsilon_{\rm CP} Y_N(T_0) = \frac{3}{4} \epsilon_{\rm CP}
\delta \left(1 + r_c - \delta \frac{M}{T_c} \right)^{-3/4} \,. 
\end{equation}
Given that $\d\approx 10^{-2}$ and
$\e_{\rm CP}\sim 10^{-6}$ -- $10^{-8}$,
it is possible to obtain the observed value for the baryon asymmetry
for suitable choices of the parameters. However, unlike in the previous case,
the result does depend on the parameters of the small-instanton transition,
in particular on the parameter $r_c$ which measures the brane impact.
Specifically, for large impact, $r_c \gg 1$, we have
\begin{equation}
Y_{L} \simeq \epsilon_{\rm CP}\d r_c^{-3/4} \,,
\label{yl1}
\end{equation} 
while for small impact, $r_c\ll 1$, we have instead 
\begin{equation}
Y_{L} \simeq \frac{3}{4} \epsilon_{\rm CP}\d \left(1- \delta
\frac{M}{T_c} \right)^{-3/4} \approx \epsilon_{\rm CP}\d \,.
\label{yl2}
\end{equation}

The $T_0 < M$ case can have another big advantage besides the ability to work analytically and the possibility of having information about the collision encoded in the baryon asymmetry. If $T_0$ is sufficiently low then one could envision our scenario taking place ${\it after}$ some low scale inflationary mechanism, for example, which could be used to solve the gravitino problem. In other words we have the option of choosing a low reheat temperature, with all the benefits that brings, and still being able to produce baryons. %%%%%%%%%%%%%%%%%%%%%%%%%%%%%%%%%%%%%%%%%%%%%%%%%%%%%%%%%%%%%%%%%%%%%%%%%%%%%

\section{Numerical computation of the baryon asymmetry}

If $T_0\gg M$ or $T_0 \sim M$ then lepton number violating scatterings,
which can reduce the generated lepton asymmetry by several orders of
magnitude, can no longer be neglected, and one has to solve the full
network of Boltzmann equations. In this case the expected asymmetry
from Eq.~(\ref{yls}) will be reduced by a washout factor $\kappa$, that is,
the generated asymmetry will be
\begin{equation} \label{yls2}
Y_{L} = \kappa\epsilon_{\rm CP} Y_N(T_0) \frac{S_0}{S_f} \;.
\end{equation}
This has been studied previously in the 
standard scenario of thermal leptogenesis \cite{PluemiDiss}. 
A characteristic feature 
of this scenario is that the generated baryon asymmetry mostly on the mass parameter
\begin{equation}
  \tilde{m}_1={(h_\nu h_\nu^\dagger)_{11}\over M_1}v_2^2,
\end{equation}
where $v_2$ is the vacuum expectation value of the MSSM Higgs field
which gives Dirac masses to up-type quarks and neutrinos. This is
due to the fact that all the scattering and decay terms entering
in the Boltzmann equations are proportional to some power of
$\tilde{m}_1$. Thermal
leptogenesis is only possible in a rather narrow range of $\tilde{m}_1$.
If $\tilde{m}_1$ is too low, the Yukawa interactions are too weak
to produce a sufficient number of RHNs at high temperatures, whereas
for large $\tilde{m}_1$ the lepton number violating scattering processes
mediated by the RHNs are too strong and destroy any generated asymmetry
\cite{PluemiDiss}.
\begin{figure}
        \centerline{\epsfig{file=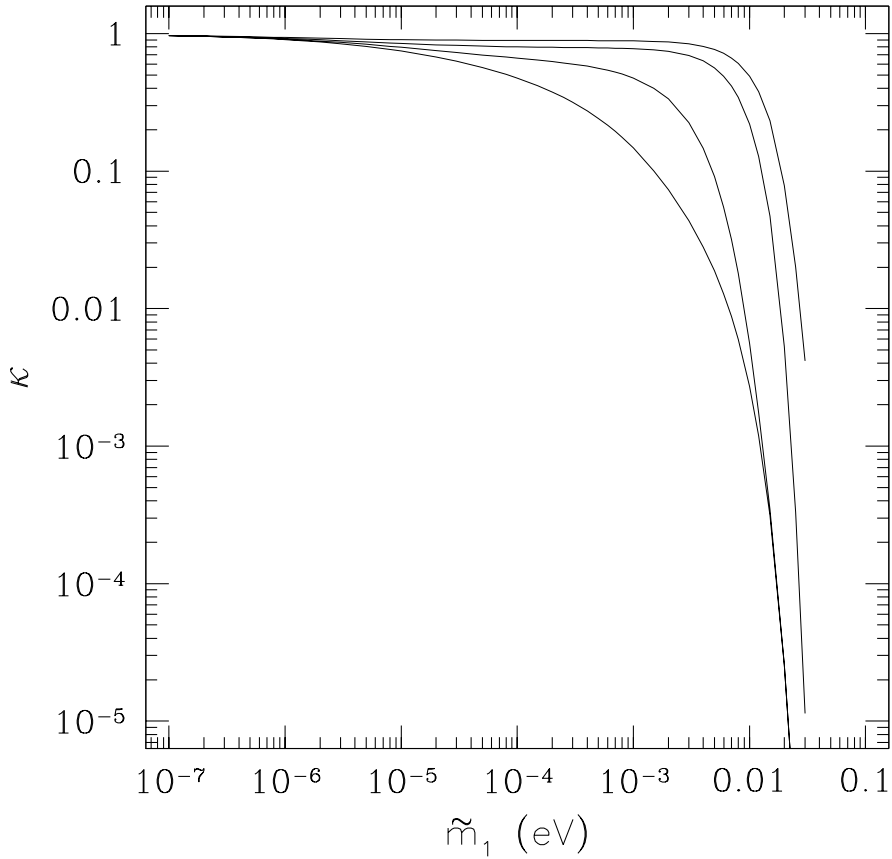,width=12cm}}
        \caption{\emph{The washout parameter $\kappa$ as a function
        of $\tilde{m}_1$ for different values of the initial temperature
        $\frac{M}{T_0}=0.1$, $5$, $10$ and $20$ (from left to right).}}
        \label{fig2}
\end{figure}

In order to see whether this is also the case in our scenario we
have numerically solved the set of Boltzmann equations, using 
initial conditions as discussed in section 3 and starting
the simulation at different values of $T_0$. The results are shown
in Fig~(\ref{fig2}), where we have plotted the washout factor 
$\kappa$ as a function of $\tilde{m}_1$, for initial temperatures 
$\frac{M}{T_0}=0.1$, $5$, $10$ and $20$. Further, we have assumed a hierarchy
of RHN masses of the form $M_1=10^{10}\;$GeV, $M_2=3\times10^{11}\;$GeV
and $M_3=10^{13}\;$GeV.

Fig.~\ref{fig2} shows that $\kappa$ converges towards unity for small
$\tilde{m}_1$, independently
of the starting temperature $T_0$, since then the washout processes
are suppressed and are out of equilibrium when the lepton asymmetry
is produced. For larger $\tilde{m}_1$ the generated asymmetry starts
to depend on $T_0$, since then the lepton number violating scattering
processes can still be in thermal equilibrium at temperatures below
the RHN neutrino mass, that is, even for $T_0<M$ the generated asymmetry
can be reduced by washout processes. Eventually however, the washout
processes will freeze out, i.e.\ a smaller $T_0$ will result in a
larger remaining asymmetry. If $\tilde{m}_1$ gets very large,
above $\sim3\times10^{-2}\;$eV, the washout processes remain in
thermal equilibrium down to very low temperatures, hence even
for $T_0\ll M$ the generated lepton asymmetry is strongly suppressed.
In summary, we see that $\kappa$ approaches one for an increasing range in
$\tilde{m}_1$ as $T_0$ decreases, which justifies our estimate for the
$T_0\ll M$ case in the previous section.

The discussion of this section should make one fact about this scenario clear; for certain possible parameter ranges in the MSSM our scenario is distinguishable experimentally from more conventional instances of leptogenesis. If at future accelerator experiments, for example, the parameters of the MSSM were measured then they could be found to take values such that the conventional picture of leptogenesis is not phenomenologically viable due to problems with washout. In such a situation our scenario would not  be ruled out  and the conventional scenario would be - the two would have been distinguished by experiment. %%%%%%%%%%%%%%%%%%%%%%%%%%%%%%%%%%%%%%%%%%%%%%%%%%%%%%%%%%%%%%%%%%%%%%%%%%%%%

\section{Conclusions and outlook}

Let us summarise the most important points and conclusions of this
paper. We have proposed, in the context of heterotic M-theory
brane-world models, a scenario for baryogenesis based on a
small-instanton phase transition induced by a brane collision. Our scenario has a
crucial difference to more standard scenarios, such as leptogenesis.
It utilises the decoupling of the temperature after the phase transition ($T_0$)
from the RHN mass, which is a phenomenon, seen in small instanton transitions, which is qualitatively new to cosmology. This allows, for example, the generation of a
lepton asymmetry at temperatures significantly below the RHN mass. This could be of significant help in dealing with things like the gravitino problem, as mentioned at the end of section \ref{est}. We have found that, in some cases, the generated baryon asymmetry depends
on characteristics of the brane-collision, such as the ``impact'' of
the colliding brane. Most importantly, we have demonstrated that an
acceptable value for the baryon asymmetry can be obtained under
reasonable assumptions for the parameters in the model.

We have also performed a more detailed analytical as well as numerical
analysis of our scenario, the latter based on the full set of
Boltzmann equations. We have seen that the results crucially depend on
the value $M/T_0$ (where $M$ is the RHN mass and $T_0$ is the
temperature after brane collision) and the impact parameter $r_c$.
For $T_0> M$ the RHNs thermalise after the collision and before
decaying. At the same time, scattering effects may be important and may wash
out the baryon asymmetry. This case is, in fact, similar to standard
leptogenesis as can be seen from the numerical results. If, on the other hand,
$T_0 < M$ scattering effects are expected to become less important,
which is indeed what our numerical results show. This allowed us to
perform an analytic estimate in this case. It turns out that, between,
roughly, $O(M/50)<T_0<O(M)$ the universe is radiation-dominated after the
collision. The baryon asymmetry then depends on the parameters of the
collision such as the precise value of $T_0$ and the impact parameter
$r_c$. If $T_0<O(M/50)$, on the other hand, the energy density after
collision is dominated by massive RHNs and, hence, the universe is
matter-dominated. The baryon asymmetry is then diluted by significant
entropy generation and becomes independent of the parameters
$T_0$ and $r_c$.

Our description of the small-instanton transition was based on
a number of ``phenomenological'' assumptions. It would clearly
be desirable to carry out a more microscopical analysis.
Unfortunately, a low-energy effective description of the transition,
suitable for our purpose, is not available at present.
Developing such a description and applying it to our proposal
is an interesting challenge for future research. 

%%%%%%%%%%%%%%%%%%%%%%%%%%%%%%%%%%%%%%%%%%%%%%%%%%%%%%%%%%%%%%%%%%%%%%%%

\vspace{1cm}

\noindent
{\large\bf Acknowledgements}\\
M.~P.~would like to thank W.~Buchm\"uller and T.~Prokopec for helpful
discussions. A.~L.~is supported by a PPARC Advanced Fellowship,
M.~P.~by the EU network ``Supersymmetry and the Early Universe''
under contract number HPRN-CT-2000-00152 and J.~G.~by a Sir James Knott
Fellowship.

%%%%%%%%%%%%%%%%%%%%%%%%%%%%%%%%%%%%%%%%%%%%%%%%%%%%%%%%%%%%%%%%%%%%%%%%%

\vspace{1cm}

\appendix

\section{Perfect fluid cosmology with a moving brane}

\label{app}

In this appendix, we present the moving-brane cosmological solutions
to the action~\eqref{S4} in the presence of a perfect fluid with
equation of state $p_{\rm fluid}=w\r_{\rm fluid}$ where $w<1$ is a
constant. These solutions may have a wide range of applications
in the context of moving-brane cosmologies. For example, for
$w=-1$ they describe the motion of a brane in the background of
a cosmological constant, a result relevant to the inflationary
scenario of Ref.~\cite{Dvali:1999pa}.
For the purpose of this paper, the solutions will be used to analyse
the first stage of our baryogenesis scenario.

\vspace{0.4cm}

We start with the Ansatz
\bea
 ds^2 &=& -e^{2\n (\t)}d\t^2 + e^{2\a (\t )}d{\bf x}^2 \\
 \b &=& \b (\t) \\
 \f &=& \f (\t) \\
 z &=& z(\t )
\eea
where we have chosen flat spatial sections, for simplicity.
The energy density and pressure of the perfect fluid can be
written as
\bea
 \r_{\rm fluid} &=& \r_0 e^{-3(1+w)\a} \\
 p_{\rm fluid} &=& w\r_{\rm fluid}
\eea
where $\r_0$ is a constant. We can integrate the equation of motion
for $z$ derived from the action~\eqref{S4} to obtain
\begin{equation}
 \dot{z}= ue^{-3\a +\n -\b +\f}\; . \label{z}
\end{equation}
This result can be used to eliminate $z$ so that we remain with a
closed set of equations for the fields $\a$, $\b$ and $\f$. Using
the formalism of Ref.~\cite{Lukas:1997iq}, these equations can be very
elegantly summarised by an effective ``moduli space'' Lagrangian given by
\be
 {\cal L} = \frac{1}{2}E{\bal '}^TG\bal '-E^{-1}U \label{dan_lag}\; .
 \label{L}
\ee
Here we have arranged the fields into a vector $\bal = (\a ,\b ,\f )^T$ 
and $G={\rm diag}(-3,\frac{3}{4},\frac{1}{4})$ is a constant metric.
The ``Einbein'' $E$ is a non-dynamical field defined by
$E=e^{3\a -\n}$ whose equation of motion leads to the Friedmann equation.
Finally, the potential $U$ on the moduli space has the structure
\begin{equation}
 U = \frac{1}{2}\sum_{r=0}^1u_r^2 e^{{\bf q}_r\cdot \bal}
\end{equation}
The first term originates from the perfect fluid where
\begin{equation}
 u_0^2=\r_0\; ,\qquad {\bf q}_0 = (3(1-w),0,0)^T
\end{equation}
while the second one is due to the moving brane where
\begin{equation}
 u_1^2 = \frac{1}{2}q_5u^2\; ,\qquad {\bf q}_1 = (0,-1,1)^T\; .
\end{equation}
The vectors ${\bf q}_r$ are characteristic for the respective origin of
the potential terms. Note that in our particular case we have
\begin{equation}
 <{\bf q}_0,{\bf q}_1>\equiv {\bf q}_0^TG^{-1}{\bf q}_1=0
\end{equation}
This implies that we are dealing with an $SU(2)^2$ Toda model which
can be integrated exactly. Following Ref.~\cite{Lukas:1997iq}, one can find
the general solution in the gauge $E=1$, that is, $\n = 3\a$.
It is given by
\bea
 \a &=& \frac{1}{\sqrt{3}}\s_0 \label{alsol}\\
 \b &=& \frac{1}{\sqrt{3}}\s_1+\s_2\\
 \f &=& -\sqrt{3}\s_1+\s_2\\
 \n &=& 3\a
\eea
with the modes
\bea
 \s_0 &=& -q_0^{-1}\ln\left[\frac{u_0^2}{k_0^2}\sinh^2(y_0)
          \right] \\
 \s_1 &=& q_1^{-1}\ln\left[\frac{u_1^2}{k_1^2}\cosh^2(y_1)
          \right] \\
 \s_2 &=& k_2(|\t |-\t_2)
\eea
and 
\begin{equation}
 y_r = \frac{1}{2}|k_r|q_r(|\t |-\t_r)
\end{equation}
where $r=0,1$. Further, the constants $q_r$ are defined by
\begin{equation}
 q_0 = \sqrt{3}(1-w)\; ,\qquad q_1 = \frac{4}{\sqrt{3}}\; ,
\end{equation}
and $\t_i$, where $i=0,1,2$, are arbitrary integration constants.
The integration constants $k_i$, where $i=0,1,2$, are subject 
to the constraint
\begin{equation}
 k_1^2+k_2^2=k_0^2\label{cons}
\end{equation}
which originates from the Friedmann equation. Inserting these results
into Eq.~\eqref{z} we find that the brane motion is described by
\begin{equation}
 z = z_0+\frac{d}{2}\tanh (y_1)
\end{equation}
where $z_0$ is an arbitrary constant and the maximal distance $d$ by
which the brane moves is given by
\begin{equation}
 d = \frac{2\sqrt{3}|k_1|}{q_5u}\; .
\end{equation}

\vspace{0.4cm}

Let us discuss some properties of these solutions. It is straightforward
to show that the total kinetic energy density $\r_{\rm kin}$ and the
energy density of the fluid $\r_{\rm fluid}$ for our solutions are
given by
\bea
 e^{2\n} \r_{\rm kin} &\equiv&\frac{1}{4}\dot{\f}^2+\frac{3}{4}\dot{\b}^2
        +\frac{q_5}{2}e^{\b -\f}\dot{z}^2 = k_1^2+k_2^2\label{kin}\\
 e^{2\n} \r_{\rm fluid} &\equiv& e^{2\n}\r = \frac{k_0^2}{\sinh^2(y_0)}\; .
 \label{fluid}
\eea
Defining the ratio $r$ of kinetic to fluid energy density one then finds
in view of Eq.~\eqref{cons}
\begin{equation}
 r\equiv\frac{\r_{\rm kin}}{\r_{\rm fluid}}=\sinh^2(y_0)\; .
\end{equation}
{}From the solution for $\a$, Eq.~\eqref{alsol}, this leads to the simple
scaling law
\begin{equation}
 \frac{a(\t_1)}{a(\t_2)} = \left[\frac{r(\t_1)}{r(\t_2 )}
                           \right]^{-1/(3(1-w))}\; .
\end{equation}
where $a=e^\a$ is the scale factor.

For a gas with temperature $T$ we have $T\sim a^{-1}$ (assuming
isentropic evolution and unchanged number of degrees of freedom)
which, using $w=\frac{1}{3}$, implies
\begin{equation}
 \frac{T_1}{T_2}=\left[\frac{r(T_1)}{r(T_2)}\right]^{1/2}\; .
\end{equation}

%%%%%%%%%%%%%%%%%%%%%%%%%%%%%%%%%%%%%%%%%%%%%%%%%%%%%%%%%%%%%%%%%%%%%%%%%%%

%%%%%%%%%%%%%%%%%%%%%%%%%%%%%%%%%%%%%%%%%%%%%%%%%%%%%%%%%%%%%%%%%%%%%%%%%
%%%%%

\end{document}